\documentclass[twocolumn,showpacs,prl,aps]{revtex4}

\usepackage{epsfig}
\usepackage{amssymb,amsmath,amsthm}

\begin{document}
\title{Spatiotemporal pulses in a liquid crystal optical oscillator}

\author{U. Bortolozzo$^1$, A. Montina$^2$, F.T. Arecchi$^2$, J.P. Huignard$^3$ and S. Residori$^4$} 

\address{$^1$Laboratoire de Physique Statistique de l'ENS, 24 rue Lhomond, 75231 Paris Cedex 5, France
\\
$^2$Physics Department, University of Florence, Largo E. Fermi 6, 50125 Florence, Italy
\\
$^3$Thales Research \& Technology, RD 128 91767 Palaiseau Cedex, France
\\
$^4$Institut Non Lin\'eaire de Nice, 1361 route des Lucioles 06560 Valbonne - Sophia Antipolis France
}

\date{\today}

\begin{abstract}
A nonlinear optical medium results by the collective orientation of
liquid crystal molecules tightly coupled to a transparent photoconductive layer. We show that such a medium can give a large gain, thus, if inserted in a ring cavity, it results in an unidirectional optical oscillator.
Dynamical regimes with many interacting modes are made possible
by the wide transverse size and the high nonlinearity of the liquid crystals. We show the generation of spatiotemporal pulses, coming from the random superposition of many coexisting modes with different frequencies.
\end{abstract}


\maketitle

Optical oscillators have been extensively studied in the past, in particular ring cavities with photorefractive gain \cite{Huignard,Arecchi} have attracted much attention, both experimentally and theoretically \cite{Anderson,Dalessandro}. Here we present a different optical medium that results from the collective orientation of
liquid crystal molecules tightly coupled to a photoconductive layer and pumped in a two-wave mixing configuration \cite{Brignon,ouroptlett}. We show for the first time that, when inserted in a ring cavity, the liquid crystal light valve has a gain large enough to overcome the losses, thus resulting in an unidirectional optical oscillator. The wide transverse size and the high nonlinearity of the liquid crystal light valve allows us to explore dynamical regimes remained inaccessible up to date, where a huge number of modes are interacting. By changing the control parameters, the system displays either regular regimes, such as the alternation of Gauss-Laguerre cavity modes and the emission of out of axis large order modes, or space-time chaotic behaviors like the generation of high amplitude spatiotemporal pulses appearing in random space points.

In this letter, we focus our attention mostly on the spatiotemporal pulses. We show that they are confined along the three space directions and that they behave as coherent structures, with a characteristic size and life-time. We present a theoretical model for the liquid crystal oscillator and, for a set of parameters consistent with the experiment, we confirm the appearance of spatiotemporal pulses by numerical simulations.
It is worth, here, to distinguish these spatiotemporal pulses from the solitary or localized structures that arise when a system presents bistability between a spatially homogenous and a pattern state \cite{locstruct}. While spatiotemporal pulses originate from a coherent superposition randomly taking place within a large number of coexisting modes, localized structures are isolated cells of a corresponding spatial pattern. Moreover, spatiotemporal pulses are confined both in space and time, while localized structures are confined only in space and can persist for indefinite time if not erased by a local, and large enough, perturbation.
Spatiotemporal pulses are somehow reminiscent of the pulsating solitons predicted in the complex Ginzburg-Landau equation (CGLE) for mode-locked lasers \cite{Akhmediev} or of the random-phase solitons found in the
nonlinear Schroedinger equation (NLSE) for nonlinear lattices \cite{Segev}. In our case, due to the large time scale separation between the fast decay of 
the field in the cavity ($\tau_c \simeq 10^{-7}$ $s$) and the slow response time of the liquid crystals ($\tau \simeq 10^{-3}$ $s$), the model cannot be reduced to a single equation for the electric field, the dynamics being slaved by the slow evolution of the refractive index. Nevertheless, the model contains the main ingredients for the creation of spatiotemporal pulses, that is, a large number of modes with partially uncorrelated phases and interacting through the nonlinear medium. Note that similar mechanisms are also at the basis of the collapsing filaments predicted in optical turbulence \cite{Dyachenko}.

\begin{figure} [h!]
\centerline{\includegraphics[width=6 cm]{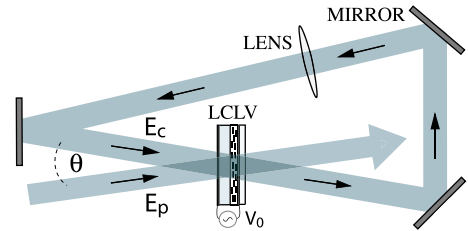}}
\caption{Experimental setup.
\label{setup}}
\end{figure} 

The experimental setup is schematically represented in Fig.\ref{setup}.
The cavity, that has a total cavity length $L=240$ $cm$, consists of three high-reflectivity dielectric mirrors and a lens of $f=70$ $cm$ focal length, which enhances the mode stability providing a nearly spherical configuration. This configuration also ensures the presence of different longitudinal modes that oscillate at different frequencies and thus can lead to the formation of spatiotemporal pulses \cite{Ponomarenko}. 
The gain medium is a liquid crystal light valve with one of the walls made of a thin slice, $1$ $mm$ thickness, of the photoconductive $B_{12}SiO_{20}$ (BSO) crystal \cite{ouroptlett}. An external voltage $V_0$ is applied to the light valve by means of transparent electrodes deposited over the glass window and the external side of the BSO crystal. The BSO acts as a transparent photoconductor, thus modulating the voltage across the liquid crystals as a function of the intensity of the light passing through the cell \cite{Brignon}. 
The working point of the liquid crystals is 
fixed at $V_0 \simeq 20$ $V$, frequency $25$ $Hz$.
The thickness of the liquid crystal layer is $d=14$ $\mu m $ and the lateral size of the cell is $20$x$30$ $mm^2$. 
The cell is pumped by an enlarged and collimated  ($10$ $mm$ diameter) beam from an 
$Ar^+$ laser ($\lambda=514$ $nm$), intensity $I_p \simeq 2$ $mW/cm^2$. 

\begin{figure} [h!]
\centerline{\includegraphics[width=8.5cm]{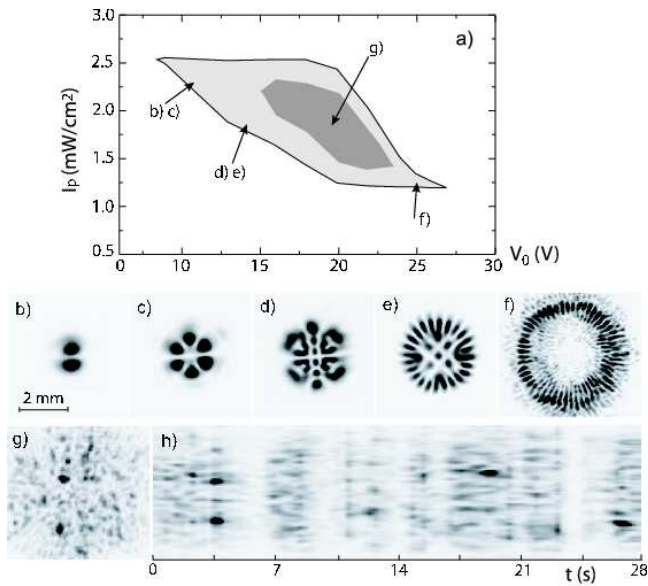}}
\caption{a) Experimental phase diagram and instantaneous snapshots taken at  b) c) $V_0=10$ $V$, $I_p=2.4$ $mW/cm^2$, d) e) $V_0=14$ $V$, $I_p=2.0$ $mW/cm^2$,
f) $V_0=25$ $V$, $I_p=1.3$ $mW/cm^2$, g)$V_0=20$ $V$, $I_p=2.0$ $mW/cm^2$; h) space-time plot obtained for the same parameter values as in g).
\label{space}}
\end{figure}

The light amplification in the cavity is based on two-wave mixing (2WM) interactions in the liquid crystals \cite{ouroptlett}. The pump beam and the cavity axis are an angle
$\theta=3$ $m rad$. The cavity field $E_c$ is spontaneously generated and has a frequency almost equal to that of the pump field $E_p$, so that the 2WM is nearly degenerate. In fact, we observe a frequency detuning of a few $Hz$, which is automatically selected by the cavity in order to maximize the gain and that corresponds to a refractive index grating slowly moving inside the liquid crystals. It is well known that a moving grating enhances the two-wave mixing gain for photorefractive crystals \cite{Huignard-book}, however this frequency shift was not observed before for a liquid crystal light valve.
Also note that the liquid crystal light valve is a thin medium, so that the 
2WM here takes place in the Raman-Nath regime \cite{Yariv}.
The Fresnel number of the cavity, which is the ratio of the area of the diffraction limiting aperture to the area of the fundamental Gaussian mode, is controlled by a diaphragm placed in front of the light valve and can be changed from $F=1$ to approximately $F=500$, which implies changing from a single transverse mode oscillation to a regime where a huge number of modes are interacting. 
An important feature, that differentiates our system from previous photorefractive cavities, is that changing the voltage $V_0$ changes the uniform refractive index $n_c$ of the liquid crystals, and thus the frequency detuning between the lowest order cavity mode and the pump beam. In order to compensate the detuning, the cavity adjusts its emission by changing the length of the out of axis transverse wave vectors \cite{Umbe}, so that the number of active modes can be changed by keeping the Fresnel number fixed and by varying the voltage $V_0$.

For the purpose of visualization, a small fraction ($4 \%$) of the cavity field is extracted by a beam sampler and directed to a CCD camera. 
A lens is inserted on the path of the extracted beam in such a way that the transverse intensity distributions recorded by the CCD are equivalent to the intensity distributions at the entrance plane of the liquid crystal light valve, $z=0$, by taking the $z$ coordinate along the cavity axis. For a fixed Fresnel number $F=500$,  and by changing  $V_0$ and the pump intensity $I_p$, we 
have determined the experimental phase diagram, as reported
in Fig.\ref{space}a. Cavity mode oscillations are in the larger grey area whereas spatiotemporal pulses are in the darker area. 
The low $V_0$ regimes are similar to those previously reported for low $F$ photorefractive cavities, with the alternation of low order Gauss-Laguerre modes \cite{Huignard,Arecchi}. The transition to the high $V_0$ regimes is accompanied by the emission of high order and out of axis symmetrical modes. 
For intermediate values of $V_0$ and $I_p$, a large number of modes interact through the nonlinear medium and give rise to the formation of spatiotemporal pulses, appearing as large intensity peaks  over a lower amplitude and "speckle-like" background. 
Instantaneous snapshots of the transverse intensity distributions $I_c(x,y)$, where $x,y$ are the coordinates in the transverse plane, are displayed in Fig.\ref{space}b-g. The transverse size of the oscillating field increases with $V_0$, up to a large ring for high $V_0$. The spatiotemporal pulses appear when a large number of modes is populating the whole size of the area illuminated by the pump beam. A typical snapshot corresponding to this case is displayed in Fig.\ref{space}g and the corresponding spatio-temporal plot is shown in Fig.\ref{space}h.

\begin{figure} [h!]
\centerline{\includegraphics[width=7.5cm]{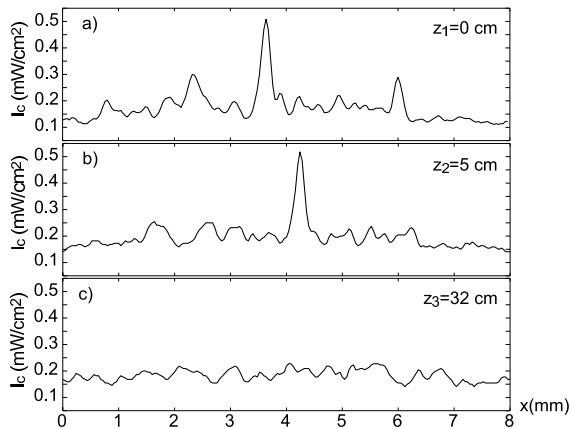}}
\caption{Spatial profiles recorded in a) $z_1$, b) $z_2$ and c) $z_3$.
\label{pulse-space}}
\end{figure}

\begin{figure} [h!]
\centerline{\includegraphics[width=7.5cm]{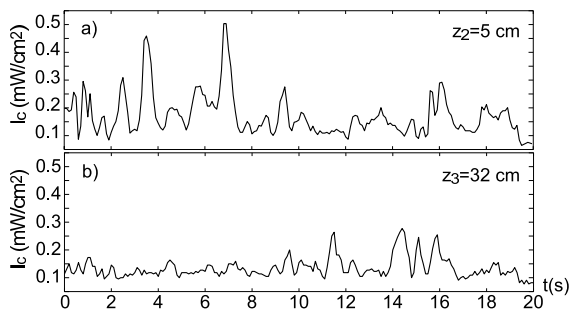}}
\caption{Temporal profiles recorded in a) $z_2$ and b) $z_3$.
\label{pulse-time}}
\end{figure}

To investigate the dynamics of spatiotemporal pulses we have fixed $V_0=20.3$ $V$ $r.m.s.$
and $I_p=2.0$ $mW/cm^2$ and we have recorded several movies. The spatiotemporal pulses are identified by applying on each frame a threshold of for instance $3$ times the average intensity, calculated as $\bar I_c=\sum_{k=1}^N I_c(x_k,y_k)/N$, with $I_c(x_k,y_k)$ the intensity of the cavity field in the $k$-th pixel point and $N$ the total number of pixels composing the image. 
The extension of the pulses in the $z$ direction is investigated in the following way: we have divided in three parts the beam extracted from the cavity, directed each part through an optical path of different length and recorded the three transverse intensity distributions by using separate CCDs.
The electronic delay time of the trigger being negligible with respect to the liquid crystal response time, we can consider that the CCDs simultaneously record the slaved field evolution in three different $z$ positions. 
By inspecting the movies, we have found that the typical longitudinal extension of the pulses is approximately $30$ $cm$, so that we have chosen to record  the intensity distributions $I_c(x,y,z_i)$ $i=1,2,3$ at distances corresponding to $z_1=0$, $z_2=5$ and $z_3=32$ $cm$.
We cut the spatio-temporal plots along the $x,y$ or $t$ direction and we obtain the 1D spatial or temporal profile of the pulses. The spatial, temporal, profiles of a pulse are displayed in Fig.\ref{pulse-space}, \ref{pulse-time}, respectively.
It can be seen from Fig.\ref{pulse-space}c and Fig.\ref{pulse-time}b that at the distance $z_3$ the pulse has disappeared.
By taking the half height width of the pulses with $I_c(x,y,z) > 3 \bar I_c$ and by averaging over more than one hundred profiles, we find that the typical transverse size of a pulse is $250 \pm 50$ $\mu m$ whereas its average lifetime is around $0.5 \pm 0.1$ $s$.

The model, derived by coupling the Maxwell equations for the cavity field with a Debye relaxation equation for the refractive index  \cite{Newell}, takes into
account the Kerr nonlinearity of the medium as well as the two-wave-mixing mechanism of photon injection inside the cavity.
The liquid crystal light valve is positioned in $z=0$, perpendicularly to the cavity axis and
$\vec r_\perp$ denotes the coordinates in the transverse $(x,y)$ plane. The refractive index $n(r_\perp,t)$ satisfies the equation
$\tau\partial_t n=n_c-n+l_0^2\nabla_\perp^2 n-\alpha |E(\vec r_\perp,t)|^2,$
where $\alpha>0$ is the nonlinear coefficient of the light valve, $n_{c}$ is a constant value determined by the voltage $V_0$,
$E(\vec r_\perp,t)=E_p e^{i(\vec k_P \cdot \vec r-\omega_p t)}+E_{c}e^{i (k_c z -\omega_p t)}+c.c.$ is the total electric field, with
$E_p$ the pump amplitude and $E_c=E_c(\vec r_\perp,t)$ the complex amplitude of the cavity field. The intensity,
$|E|^2=|E_p|^2+|E_c|^2
+(E_p^* E_c/2) e^{-i\vec k_\perp \cdot \vec r_\perp} +c.c.$,
gives rise to a refractive index change with two components, one varying slowly in space, the other corresponding to a spatial grating with wave number $\vec k_\perp=\vec k_p- \vec k_c$. We thus write the refractive index  as $n=n_{c}-\alpha|E_p|^2+n_0+(n_1/2) e^{-i\vec k_\perp\vec r_\perp}+c.c.$. By substituting this expression in the equation for $n$ we obtain 

\begin{eqnarray}
\label{enne1}
\tau {\partial n_0 \over \partial t}&=&(-1+l_0^2\nabla_\perp^2)n_0-\alpha |E_c|^2
\\
\nonumber
\tau {\partial n_1\over \partial t}&=&\left(-1+l_0^2\nabla_\perp^2 +l_0^2 |\vec k_\perp|^2\right)n_1-2 i l_0^2\vec k_\perp\cdot\vec\nabla n_1+
\\
\nonumber
&-&\alpha E_p^*E_c
\end{eqnarray}
for the slowly varying fields $n_0$ e $n_1$.
The wave equation for the cavity field is written by adopting, as usual, the slowly varying amplitude approximation and by considering a planar cavity. Due to the large scale separation between the medium response time  and the cavity round-trip time, we neglect the time derivative and obtain

\begin{eqnarray}
\label{statio_plan}
\frac{\partial E_c}{\partial z}&=&\left[\frac{i}{2 k_c}
\nabla_\perp^2+i k_c n_0W(z)
+\frac{i\delta-\gamma_c}{c}\right]E_c+\\
\nonumber
&+&i k_c n_1 W(z)E_p,
\end{eqnarray}
where $W(z)=1$ in the liquid crystal light valve, $z=0$ to $d$, and $0$ elsewhere, 
$\delta$ is the frequency detuning between the lowest order cavity mode and the pump field and
$\gamma_c$ is the cavity loss rate. When deriving this equation we have considered $n_0$ and $n_1$ small.

Inside the liquid crystal light valve we neglect the transverse Laplacian and we integrate in $z$ the field equation, Eq.(\ref{statio_plan}), thus obtaining

\begin{equation}
E_c(d) = e^{i k_c d n_0} \left [ J_0(2k_c d | n_1|) E_c(0)+i J_1(2k_c d | n_1 |) E_p(0) \right ],
\label{Raman-Nath} 
\end{equation}
which accounts for Raman-Nath diffraction \cite{Yariv} of the pump in the 2WM process.
Outside the liquid crystals the field evolution is governed by diffraction, thus the transverse Laplacian has to be retained.
By considering that the cavity field has to satisfy the periodic boundary conditions imposed by the cavity, and by taking into account the presence of the lens, it can be shown \cite{noi} that the field at the entrance of the cell is given by

\begin{equation}
E_c=i \sum_{m=1}^\infty \hat B^m \hat C E_p(0),
\label{field-equation}
\end{equation}
where

\begin{equation}
\hat B =e^{i\frac{ L_2}{2k_c}\nabla_\perp^2}e^{-i\frac{k_c}{2f} \vec r_\perp^2}
e^{i\frac{L_1}{2k_c}\nabla_\perp^2 +\frac{i\delta-\gamma_c}{c} L}e^{id k n_0} J_0(2k_c d | n_1|) 
\label{B}
\end{equation}
and

\begin{equation}
\hat C = e^{i\frac{ L_2}{2k_c}\nabla_\perp^2}e^{-i\frac{k_c}{2f} \vec r_\perp^2}
e^{i\frac{L_1}{2k_c}\nabla_\perp^2 +\frac{i\delta-\gamma_c}{c} L}e^{id k n_0} J_1(2k_c d | n_1|),
\label{C}
\end{equation}
with $L_1$ the distance between the liquid crystal light valve and the lens, $L_2=L-L_1$.
The interpretation of the above equations is quite simple. The first term of the sum, $E_c^{(0)}=i \hat C E_p(0)$, is Eq.(\ref{Raman-Nath}) with $E_c(0)=0$ plus the field evolution in the cavity, thus accounts for the field generated by the pump in one cavity round-trip. The following terms sum up the field evolutions on the successive cavity loops.
Inside the liquid crystal light valve, the operator $\hat B$ gives the same field equation as 
Eq.(\ref{Raman-Nath}) but with $E_p(0)=0$.
At variance with usual treatments, where the mean-field approximation is used to eliminate the $z$ dependence of the field, our model keeps this dependence, which is necessary to accounts for the formation of 3D patterns. Previously, the $z$ behavior was considered for optical parametric oscillators \cite{Leberre} and in dispersive self-focusing media \cite{Silberberg}, where optical bullets have been predicted as the 3D analogous of optical solitons.

\begin{figure} [h!]
\centerline{\includegraphics[width=7.5cm]{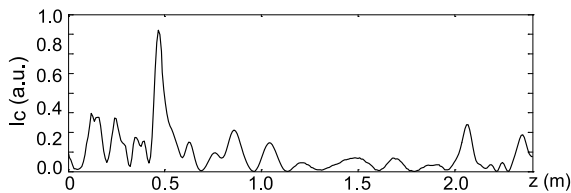}}
\caption{Numerical pulse profile along $z$.
\label{profiles-num}}
\end{figure}

\begin{figure} [h!]
\centerline{\includegraphics[width=8cm]{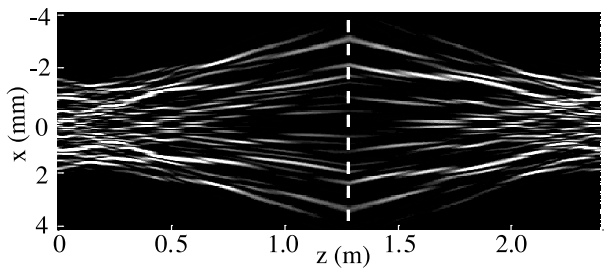}}
\caption{Numerically calculated field distribution in the $x-z$ plane at a fixed time; the dashed line marks the lens position ($L_1=1.3$ $m$, $L_2=1.1$ $m$ and $f=0.7$ $m$).
\label{numerics}}
\end{figure}

We have performed numerical simulations of the model equations, Eq.(\ref{enne1}) and Eq.(\ref{field-equation}) in the 2D+1 approximation, that is, by keeping the time and the $x$,$z$ spatial dependences. The parameters are chosen from the typical values of the experiment, $\tau=40$ $ms$, $l_0=30$ $\mu m$,$I_p=2$ $mW / cm^2$ and $\alpha=4$ $cm^2/W$. The number of terms in the sum is truncated to the number of round-trips given by the average lifetime of photons in the cavity.
In Fig.\ref{profiles-num} is shown the profile of a spatiotemporal pulse along $z$. In Fig.\ref{numerics} we show the intensity distribution calculated in the $x,z$ plane at a fixed time. In agreement with the experiment, spatiotemporal pulses appears as filaments confined both in the transverse direction and along the light propagation direction $z$.

In conclusion, we have shown a new type
of nonlinear optical oscillator, which includes a thin liquid crystal light valve as the gain medium, and we have given evidence of the formation of spatiotemporal pulses.  

U. Bortolozzo acknowledges a fellowship of the {\it Ville de Paris}. A. Montina acknowledges a fellowship of the {\it Ente Cassa di Risparmio di Firenze}, under the project "dinamiche cerebrali caotiche".

\end{document}